\documentclass{article}

\usepackage{arxiv}

\usepackage[utf8]{inputenc} 
\usepackage[T1]{fontenc}    
\usepackage{hyperref}       
\usepackage{url}            
\usepackage{booktabs}       
\usepackage{amsfonts}       
\usepackage{nicefrac}       
\usepackage{microtype}      
\usepackage{lipsum}
\usepackage{graphicx}
\usepackage{multirow}
\usepackage{caption}
\usepackage{amsmath,amsfonts}
\usepackage{algorithmic}
\usepackage{algorithm}
\usepackage{array}
\usepackage[caption=false,font=normalsize,labelfont=sf,textfont=sf]{subfig}
\usepackage{textcomp}
\usepackage{stfloats}
\usepackage{url}
\usepackage{verbatim}
\usepackage{graphicx}
\usepackage{cite}
\usepackage{subcaption}
\usepackage{xcolor}
\usepackage{multirow}

\def \etal {et al. }
\graphicspath{ {./images/} }

\title{Describe Where You Are: Improving Noise-Robustness for Speech Emotion Recognition with Text Description of the Environment}

\author{
 Seong-Gyun Leem \\
  Department of Electrical and Computer Engineering \\
  The University of Texas at Dallas\\
  Richardson, TX 75080 USA \\
  \texttt{SeongGyun.Leem@utdallas.edu} \\
   \And
 Daniel Fulford \\
  Occupational Therapy and \\
  Psychological and Brain Sciences \\ 
  Boston University\\ 
  MA 02215 USA\\
  \texttt{dfulford@bu.edu} \\
  \And
  Jukka-Pekka Onnela \\
  Department of Biostatistics, \\
  Harvard T.H. Chan School of Public Health\\
  Harvard University \\
  MA 02138 USA \\
  \texttt{onnela@hsph.harvard.edu} \\
  \AND
  David Gard \\
  Psychology Department \\
  San Francisco State University \\
  CA 94132 USA \\
  \textit{dgard@sfsu.edu} \\
  \And
  Carlos Busso \\
  Department of Electrical and Computer Engineering \\
  The University of Texas at Dallas\\
  Richardson, TX 75080 USA \\
  \texttt{busso@utdallas.edu} \\
}

\begin{document}
\maketitle
\begin{abstract}
\emph{Speech emotion recognition} (SER) systems often struggle in real-world environments, where ambient noise severely degrades their performance. This paper explores a novel approach that exploits prior knowledge of testing environments to maximize SER performance under noisy conditions. To address this task, we propose a text-guided, environment-aware training where an SER model is trained with contaminated speech samples and their paired noise description. We use a pre-trained text encoder to extract the text-based environment embedding and then fuse it to a transformer-based SER model during training and inference. We demonstrate the effectiveness of our approach through our experiment with the MSP-Podcast corpus and real-world additive noise samples collected from the Freesound and DEMAND repositories. Our experiment indicates that the text-based environment descriptions processed by a \emph{large language model} (LLM) produce representations that improve the noise-robustness of the SER system. With a \emph{contrastive learning} (CL)-based representation, our proposed method can be improved by jointly fine-tuning the text encoder with the emotion recognition model. Under the -5dB \emph{signal-to-noise ratio} (SNR) level, fine-tuning the text encoder improves our CL-based representation method by 76.4\% (arousal), 100.0\% (dominance), and 27.7\% (valence).
\end{abstract}


\section{Introduction}
\label{sec:intro}
 \emph{Speech emotion recogntion} (SER) systems have highly improved with the help of pre-trained speech representation models \cite{Baevski_2020, Hsu_2021, Chen_2022} and the creation of larger emotional speech databases \cite{Martinez-Lucas_2020, Kondratenko_2023, Lotfian_2019_3,Upadhyay_2023_2}. Recently, there has been increased interest in deploying SER systems in real-world applications, opening opportunities across many domains,  such as digital assistants \cite{Chatterjee_2021}, health care applications \cite{Fulford_2021}, and security and defense. One important barrier in this direction is the degradation of SER performance in real-world environments caused by multiple types of non-stationary background noise \cite{Lee_2021}. 

Several solutions have been proposed to improve the robustness of SER systems against acoustic noise. The solutions include data augmentation \cite{lakomkin_2018, Wu_2023_2, Ranjan_2024}, feature enhancement \cite{juszkiewicz_2014, Triantafyllopoulos_2019}, feature selection \cite{Schuller_2006, Leem_2022}, and domain adaptation approaches \cite{Wilf_2021, Leem_2021}. Since transformer-based speech representation models have been successfully used in speech problems \cite{Baevski_2020, Hsu_2021, Chen_2022}, many studies have also worked on increasing the noise robustness of SER systems built with pre-trained speech representation models \cite{Mitra_2023, Leem_2023}. These approaches can increase the performance of transformer-based SER models in target noisy conditions. However, it is challenging to use these models in scenarios with multiple noisy environments since a transformer-based SER model requires important resources to adapt and store its parameters for each target environment. To address multiple noise types in a single SER model, Leem \etal \cite{Leem_2023_2} proposed environment-agnostic and -specific adapters. Their work showed that leveraging the prior knowledge of the testing condition is important for an SER model's adaptation to multiple noisy environments.

This paper explores which form of prior knowledge allows an SER model to effectively adapt to multiple unseen environments. Rather than aiming to cover all the environments, our system trained the model to be conditioned by a text embedding describing the environment, which  project the unseen condition into the ones that are the closest to the seen environments. With this strategy, the prior knowledge is used as a mechanism for zero-shot learning in new environments with types of noises not considered while training the models. It also provides the mechanism to indirectly identify similar environmental conditions during training (e.g., noise in a bus station and a train station). Exploring this problem, we investigate using text-based environment descriptions as the prior knowledge for a noise-robust SER system. Using natural language prompts during training has shown potential in image classification \cite{Radford_2021}, sound event classification \cite{Wu_2023}, and several speech processing downstream tasks, including keyword spotting, and speaker counting \cite{Elizalde_2023}. Natural language supervision is also applicable to SER tasks \cite{Stanley_2023, Gong_2023}. All these studies indicate that exploiting text information is a promising strategy for SER systems. We propose a \emph{text-guided environment-aware training} (TG-EAT) strategy to improve the noise robustness of an SER model with text descriptions. We focus on the prediction of arousal (calm to active), valence (negative to positive), and dominance (weak to strong). TG-EAT uses noisy speech and its text-based environmental description to adapt the SER model. We use a pre-trained text encoder to extract the representation of text-based environment descriptions. This representation is combined with a transformer-based SER model. During adaptation, the SER model learns appropriate denoising functions with respect to the given environment description. During inference, we only need to change the template sentence to guide the SER model with testing environment information. We expect that the pre-trained text encoder can capture similar semantic information from environmental conditions included in the train set, allowing zero-shot environment learning for the SER model. This approach is expected to generalize the SER performance when tested in environmental conditions that are not included in the training process. 

Our experiment with the MSP-Podcast corpus shows that using text descriptions of the testing environment can highly improve the SER performance, especially with \emph{large language model} (LLM). In the -5 dB \emph{signal-to-noise ratio} (SNR) condition, our method improves the original SER model built with a \emph{self-supervised learning} (SSL) representation by 7.6\% for arousal, 8.3\% for dominance, and 45.4\% for valence. When we compare the proposed SER model with the DAT baseline, we observe improvements of 16.6\% for arousal, 18.1\% for dominance, and 23.0\% for valence (-5 dB SNR level). With the text encoder from CLAP, pre-trained with paired audio, the SER model can achieve the best performance in the low SNR condition. Compared to freezing the text encoder, the fine-tuning approach improves performance by 76.4\% for arousal, 100.0\% for dominance, and 27.7\% for valence under the -5 dB SNR condition. Our solution is highly applicable to SER systems deployed in real-world applications. For example, systems can infer the testing environment from a \emph{global positioning system} (GPS) by using \emph{geological information service} (GIS) mashups, such as OpenStreetMap \cite{Vargas-Munoz_2021}. The main contributions of this study are:

\begin{itemize}
    \item We explore using text embedding for an SER model to increase noise robustness in unseen conditions by explicitly leveraging the environment information. Our method provides a unique advantage by enabling a single model to adapt to multiple noise conditions using text embeddings rather than requiring multiple context-specific expert models. This is particularly beneficial for transformer-based architectures, which demand significant computational resources while delivering SOTA performance for SER. By leveraging text-described target environment information, we maximize performance without the overhead of maintaining multiple models.
    \item We show the benefits of using LLM to improve SER performance under noisy conditions over using a pre-trained environment classifier, especially in a low SNR condition.
    \item We show that fine-tuning the text encoder of CLAP can improve SER performance, leading to the possibility of using a paired audio encoder to deal with unknown testing environments.
\end{itemize}

Our paper is organized as follows. Section \ref{sec:previousstudies} describes studies relevant to SER in noisy conditions and text-guided training strategies. Section \ref{sec:proposed_method} describes the proposed approach, emphasizing the motivations and insights behind the TG-EAT framework. Section \ref{sec:experimental_settings} provides the experimental setting, including the database, baselines, and implementation details. Section \ref{sec:results} presents the results, discussing the clear benefits of the proposed strategy. Finally, Section \ref{sec:conclusion} concludes the paper, summarizing our study and providing future research directions inspired by the proposed approach.

\section{Previous Work}
\label{sec:previousstudies}

\subsection{Speech Emotion Recognition under Noisy Environments}
\label{ssec:noisy_ser}

Increasing the noise robustness of an SER system is an essential task when deploying it in real-world applications. Previous studies have mainly focused on improving acoustic features for the SER model. 
Triantafyllopoulos \etal \cite{Triantafyllopoulos_2019} proposed to enhance noisy waveforms before extracting the SER features. The enhancement models used \emph{convolutional neural network} (CNN) with residual blocks. Pandharipande \etal \cite{Pandharipande_2018} proposed to discard noisy frames to increase the noise robustness of an SER model by using a voice activity detection module. Leem \etal \cite{Leem_2024} proposed to select noise-robust LLDs by addressing the performance and robustness of each single LLD. 

More recently, SER studies have mainly focused on using transformer-based speech representation models \cite{Goncalves_2024,Pepino_2021,Wagner_2023,Mote_2024, Goncalves_2024_2, Upadhyay_2024}, including Wav2Vec2.0 \cite{Baevski_2020}, HuBERT \cite{Hsu_2021}, and WavLM \cite{Chen_2022}. Such models have shown higher robustness against  small perturbations on the input speech than the traditional SER model with a Mel-spectrogram \cite{Wagner_2023}. Despite this trend, they still show performance differences from the ones tested in a clean environment. For this reason, studies are currently exploring strategies to improve the noise robustness of the pre-trained speech representation model. A common approach to address this issue is noise-aware training, where the clean training set is augmented with the noise sound during environment adaptation. Mitra \etal \cite{Mitra_2023} demonstrated that training a HuBERT-based SER model with noisy speech can highly improve the performance in low SNR conditions. Leem \etal \cite{Leem_2023} proposed a contrastive teacher-student learning strategy to address the catastrophic forgetting issue when training a fine-tuned SER model with noisy speech. Wu \etal \cite{Wu_2023_2} proposed to dynamically change the distortion level of the augmented speech during adaptation based on the distortion metrics. 

The aforementioned methods focused on increasing the SER model's robustness against a single target environment. They might not be the optimal solution for an SER model deployed on a real-world application since it is highly likely that this system will encounter multiple types of environmental noises. We focus on adapting a single transformer-based SER model to multiple noisy environments to efficiently deal with multiple types of environments. To address this issue, Leem \etal  \cite{Leem_2023_2} proposed to adapt the transformer-based SER model to multiple types of noises with skip connection adapters. They not only trained the SER model with multiple environments but also focused on leveraging the environmental information of the testing conditions to improve SER performance under noisy conditions. The results showed that using the environment-agnostic and -specific adapters with respect to the testing condition can improve the SER performance under noisy conditions. Such prior knowledge could be achieved using domain knowledge or GPS information. Their result showed that using environmental information during inference is important for a SER model to perform well under noisy conditions. This work indicates that leveraging the prior knowledge of the testing condition is also important for a noise-robust SER model, as well as training it with multiple types of noises. This is beneficial for an SER model deployed on real-world applications where the system can exploit the domain knowledge of the testing environment and the GPS information.

This paper also explores the multi-condition training approach where the fine-tuned SER model is adapted to multiple types of noise. Different from other methods, our strategy relies on a text embedding that describes the testing environment to deal with multiple unseen environments.

\subsection{Text-Guided Training}
\label{ssec:text_training}
As we discussed in Section \ref{ssec:noisy_ser}, exploiting environmental information can improve SER performance in a noisy environment. This paper mainly focuses on using text prompts to infuse environmental information into an SER model. Using natural language prompts does not require the recognition model to use a fixed set of predetermined labels during training. \emph{Contrastive language-image pre-training} (CLIP) is a good example of this approach \cite{Radford_2021}. It consists of an image encoder and a text encoder, trained with pairs of images and their corresponding text descriptions. These encoders are trained in a contrastive learning manner, which maximizes the similarity of both representations if the image and the description are paired and minimizes the similarity if they are unpaired. After training, these encoders can perform zero-shot classification by checking the similarity between the given image and the candidate prompts. The study of Radford \etal \cite{Radford_2021} used the following prompt template: \textit{``A photo of a \{label\}''}. They calculate the similarity between the representation from the given image and the representations from the prompts with different \textit{\{label\}}, selecting the \textit{\{label\}} that shows the maximum similarity. 

The contrastive pre-training strategy with natural language supervision is also successful in universal audio and speech processing. Wu \etal \cite{Wu_2023} demonstrated that pre-training audio and text encoder with natural language guidance could improve audio classification performance. The study of Elizalde \etal \cite{Elizalde_2023} showed that such natural language guidance can improve speech processing tasks, including keyword spotting, speaker counting, and SER tasks.

Previous studies have found that natural language supervision can apply to SER tasks. Stanley \etal \cite{Stanley_2023} used word embeddings to encode emotional labels for SER model. Gong \etal \cite{Gong_2023} used LLM to infer weak emotion labels for unlabeled data for weakly-supervised learning of an SER model. All these findings have shown that exploiting text information is highly applicable to SER systems. To the best knowledge of the authors, the use of natural language supervision to address SER robustness against unknown noisy environments is a novel research direction.

\section{Proposed Method}
\label{sec:proposed_method}
\begin{figure}[t]
  \centering
  \includegraphics[width=0.95\linewidth]{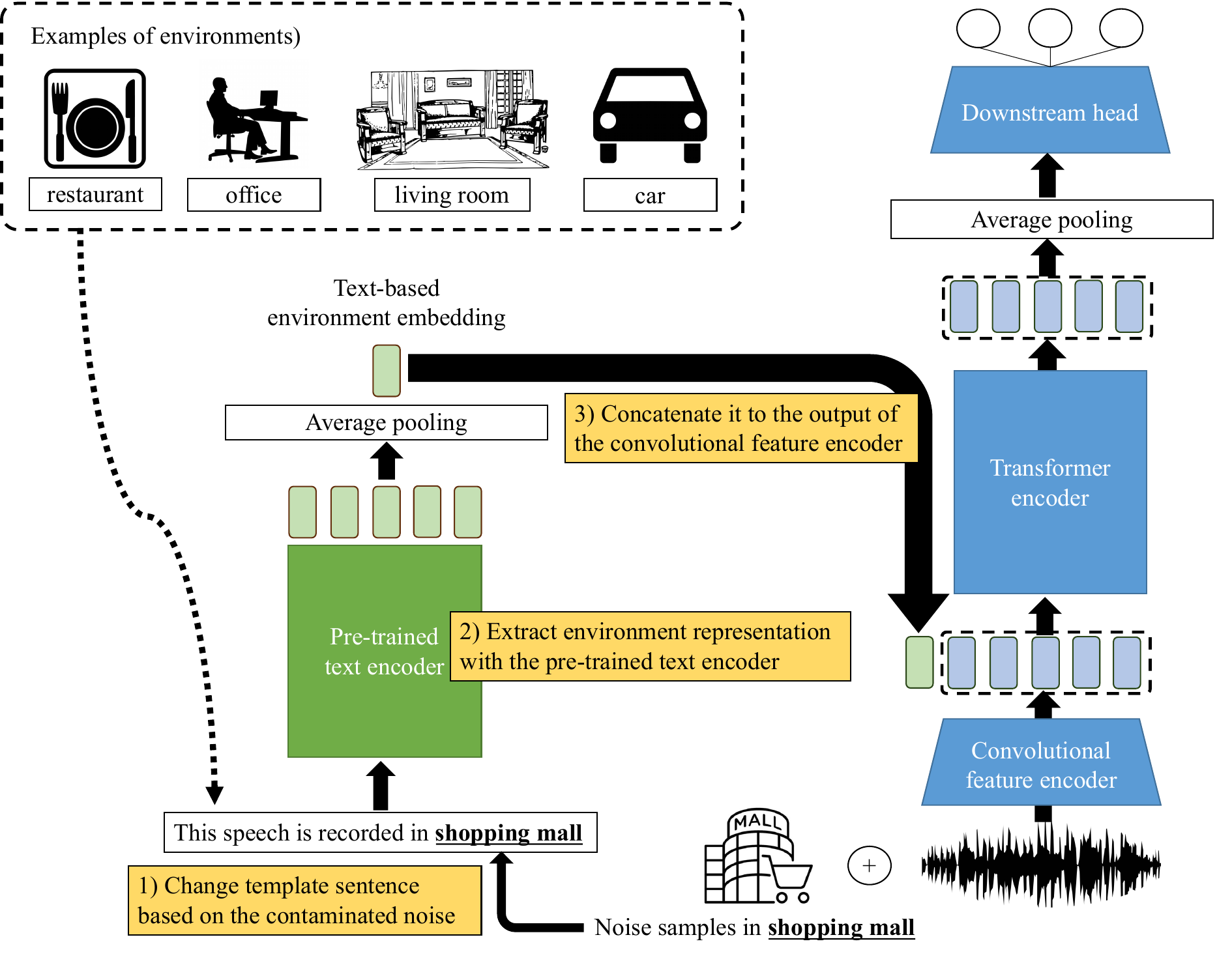}
  \caption{Our proposed text-guided environment-aware training framework. The environment representation is concatenated with the output of the convolutional feature encoder.}
  \label{fig:proposed_method}
\end{figure}

This paper proposes \emph{text-guided environment-aware training} (TG-EAT), which leverages environmental information to improve an SER model in noisy conditions. Figure~\ref{fig:proposed_method} illustrates our proposed TG-EAT framework, which uses a pair of noisy speech and its corresponding environmental description. The text embedding extracted from the environmental description is combined with the acoustic representation in the SER model, allowing it to  improve the representation for the given environmental description.

The key contribution of this study is how we use the text description from the target environment. We used prompts to generate the text description where the target environment is changed. As a preliminary experiment, we tested different prompts to describe the target environment such as \emph{``The type of background noise is \underline{\{environment\}},''} or \emph{``The input is recorded with a sound of \underline{\{environment\}}.''} We change \emph{\underline{\{environment\}}} in the prompts according to the target environment during training and testing. We found that all the prompts showed similar emotion recognition performance for all the attributes. Therefore, we consistently use the following prompt in this study: \emph{``This speech is recorded in \underline{\{environment\}}.''} We extract the text-based environment embedding from this text description using a pre-trained text encoder. We test two different text representations: \emph{contrastive learning} (CL)-based representation and LLM-based representation. For the CL-based representation, we use the text encoder pre-trained with the \emph{contrastive language audio pre-training} (CLAP) strategy \cite{Wu_2023, Elizalde_2023}. CLAP consists of an audio encoder and a text encoder. It uses a pair of acoustic events and their text description during pre-training (e.g., \emph{bird chirping sound} with the description, ``Bird is chirping in the given audio''). With these audio-text pairs, the training objective is to maximize the similarity between the audio and text representation if they are from the same pair and minimize it if they are from a different pair. Since CLAP uses an audio-text pair during pre-training, we assume that its text encoder can generate an appropriate representation from the given environment description coherent with the target acoustic condition. This paper uses the pre-trained text encoder from the unfused CLAP model proposed in the study of Wu \etal \cite{Wu_2023}. 
We take a 768-dimensional latent embedding from the text encoder, using it as our text-based environmental embedding. For the LLM-based representation, we use the encoder from the pre-trained RoBERTa model \cite{Liu_2019_2}. RoBERTa is pre-trained with \emph{masked language modeling} (MLM) and \emph{next sentence prediction} (NSP) tasks. RoBERTa has shown good performance in various benchmarks for evaluating natural language understanding systems, such as GLUE \cite{Wang_2019_4}. Although it is not pre-trained with audio data, we assume that its encoder can extract enriched semantic information from the given prompt. We use RoBERTa-large, which has 24 transformer layers. For each text encoder, we use the same tokenizer used in its pre-training to tokenize the text description of the environment.  We extract token-level text embeddings from the tokenized prompt and then apply average pooling, resulting in a 1024-dimensional single representation vector for each prompt.

After the environmental representation is obtained, the next step is to introduce this information into the model. We mainly focus on a transformer-based SER model, which has shown good performance in SER tasks \cite{Keesing_2021, Wagner_2023}. An important task is to fine-tune the model with clean and emotional speech data. We first fine-tune the SER model with clean speech to  maximize the \emph{concordance correlation coefficient} (CCC) between the predicted and the ground-truth emotional attribute scores of arousal, dominance, and valence. After fine-tuning with clean speech, the SER model is continuously updated with the training set contaminated with multiple types of noise and their corresponding text description. We insert the text representation from the given environment description into the fine-tuned transformer-based SER model. We achieve this goal by combining the text embedding with the acoustic representation, which is the output of the convolutional encoder. We apply trainable linear projection to the text embedding to match its dimension to the acoustic representation embeddings. We concatenate the projected text embedding to the acoustic representation embeddings along the time axis, then feed them into the transformer encoder.  
We choose this approach to allow the self-attention module in the transformer encoder to attend to the text embedding to all acoustic representation embeddings. Previous studies have proposed alternative approaches to add text embeddings into a prediction model \cite{Perez_2018, Alayrac_2022}, but we leave this research direction as our future investigation to further improve its performance. We update the transformer encoder and the downstream head with the concatenated embeddings. We use the same training objective as the one used when training with clean speech. From this framework, we want to evaluate if the SER model can learn the denoising function given a noisy acoustic representation with its text embedding.

\section{Experimental Settings}
\label{sec:experimental_settings}
\subsection{Data preparation}
\label{ssec:data_preparation}
Our experiment uses the MSP-Podcast corpus \cite{Busso_2025}, which consists of natural and diverse emotional speech samples from various podcast recordings \cite{Lotfian_2019_3}. The audios do not include background music or overlapped speech, and their predicted SNR is above 20 dB. We consider this corpus a clean emotion speech database for these reasons. This study focuses on predicting the emotional attributes of arousal (calm to active), dominance (weak to strong), and valence (negative to positive). Labels for these attributes were annotated by at least five raters using a seven-point Likert scale. We average the scores provided by raters for each sample to establish its ground truth values. This paper uses version 1.10 of the corpus, which consists of 104,267 annotated utterances. We use the train set to fine-tune the pre-trained speech representation model, using it as the original SER model. We use samples from the development set to select the best model during the fine-tuning process.

\begin{table}[tb] 
\caption{Keywords that are used for contaminating \emph{training}, \emph{development}, and \emph{testing} sets. \emph{Freesound} illustrates the keywords that are used for crawling the ambient recordings from the Freesound repository. \emph{DEMAND} illustrates the keywords paired with the recorded sounds in the DEMAND corpus.}
\label{tab:keywords} 

\centering 
\begin{tabular}{p{0.2\columnwidth}|l|p{0.5\columnwidth}} 
\hline 
Data Split & Corpus & Keywords \\  
\hline 
 \hline 
 Training, Development & \emph{Freesound}  & 
 mall, restaurant, office, airport, station, city, park, street, traffic, home, kitchen, living room, bathroom, bedroom, metro, bus, car, construction site, pedestrian, beach \\
 \hline
 \multirow{25}{*}{Testing}  & \emph{Freesound} &  plaza, garden, school, tram, sea, boat, amusement park, aquarium, arcade, art gallery, backyard, balcony, bank, bar, barn, beach, bridge, cafe, campground, canyon, carnival, cave, cemetery, church, circus, classroom, creek, crowd, dessert, dock, elevator, exhibition hall, factory, fairground, farmyard, festival, field, forest, fountain, gallery, gas station, grocery store, gym, harbor, highway, hospital, hotel, ice rink, industrial site, jungle, lake, laundromat, library, lobby, machine shop, market, meadow, mountain, museum, nightclub, parade, parking lot, patio, pet store, playground, pub, river, rooftop, shopping center, stadium, subway, swimming pool, theater, valley, waiting room, warehouse, waterfall, wetland, workshop, yard
 \\
 \cline{2-3}
 & \emph{DEMAND} & washroom, kitchen, living room, sports field, river, park, office, hallway, meeting, subway station, cafeteria, restaurant, traffic intersection, town square, cafe terrace, subway, bus, car
 \\
 \hline
\end{tabular} 
\end{table}

We simulate real-world noisy environments by collecting noise sounds from the Freesound repository \cite{Fonseca_2017}, which contains publicly available ambient noise sounds. We use diverse queries related to each environment to collect noise sounds, including indoor, outdoor, and in-vehicle conditions. Additionally, we included the DEMAND dataset for additional testing conditions. DEMAND contains 15 different recording conditions that simulate indoor, outdoor, and in-vehicle environments. We directly use the metadata of each recording sample to define the keyword for the testing conditions. Table \ref{tab:keywords} illustrates the keywords that are used to contaminate train, development, and test sets. We use 20 noisy environments for the train and development sets and 89 environments to contaminate the test set. Although these noise sounds are not used during adaptation, they have common characteristics with the noise sounds used during adaptation (e.g., indoor, outdoor, or in-vehicle conditions). We want to evaluate whether our proposed method can capture this semantic similarity during inference. We randomly pick the noise sounds to contaminate the Test1 set of the clean MSP-Podcast corpus. We repeat this process 10 times, creating 10 different sets for three different SNR levels, 5dB, 0dB, and -5dB. We also create a \emph{random} set, where the SNR levels are randomly selected from 5dB to -5dB. This set simulates the testing condition in a real-world application where the SNR level varies.

\subsection{Fine-Tuning Transformer-Based Architecture}
\label{sec:fine_tuning}

We implement our proposed approach with two different pre-trained speech representation models: wav2vec2-large-robust \cite{Hsu_2021_2} and the wavlm-base-plus models \cite{Chen_2022}. The wav2vec2-large-robust model has shown good performance in the emotional attribute prediction task \cite{Wagner_2023}. The wavlm-base-plus model has shown good performance for emotion recognition in the \emph{speech processing universal performance benchmark} (SUPERB) \cite{Yang_2021_3}. This model is pre-trained with noise, creating representations that are expected to be more robust to noise than other SSL representations. We fine-tune the transformer encoder of the pre-trained speech representation model and the downstream head with the clean version of the MSP-Podcast corpus. For wav2vec2-large-robust, we remove the top 12 transformer layers from the model to preserve the recognition performance with fewer parameters \cite{Wagner_2023}. We import the pre-trained models from the HuggingFace library \cite{Wolf_2019}. We use two fully connected layers for the downstream head, where each layer has 512 nodes, layer normalization, and the \emph{rectified linear unit} (ReLU) as the activation function. We use dropout in all the hidden layers to increase regularization, with a rate set to $p=0.5$. We use a linear output layer with three nodes to predict emotional attribute scores, where each node predicts the scores for arousal, dominance, and valence. We apply average pooling on top of the last transformer layer's representation to feed it to the downstream head.

During fine-tuning, we apply Z-normalization to the raw waveform by using the mean and standard deviation estimated over the training set and min-max normalization to the emotional labels, mapping them to the range of 0 to 1.  
We use the same mean and standard deviation estimated over the training set to normalize the test set's raw waveform. We use 32 utterances per mini-batch and update the model for ten epochs. We use the Adam optimizer \cite{Kingma_2014_2} with a learning rate warmup scheduling, which shows good performance when fine-tuning a pre-trained transformer architecture \cite{Popel_2018}. For the first 1,000 mini-batches, we linearly increase the learning rate from $1e^{-8}$ to $1e^{-5}$. After the 1,000 mini-batches, we fix the learning rate to $1e^{-5}$. All of our experiments are conducted on a single NVIDIA GeForce RTX 3090.

\subsection{Text-Guided Environment-Aware Training}
\label{sec:adaptation}
After fine-tuning with the clean speech, we adapt the SER model to the noisy environmental conditions. We randomly select one of the 20 noise conditions for each mini-batch during adaptation. We then use 32 different noise samples in the selected condition to contaminate 32 clean speech samples from the training set of the MSP-Podcast corpus. We build text prompts with respect to the picked environment for each mini-batch, as described in Section~\ref{sec:proposed_method}. In real-world applications, it is difficult to assume the exact SNR level of the testing condition. Therefore, we introduce an SER mismatch between our experiment's adaptation and testing stages. We randomly select the SNR level for the adaptation of the models among these options: \{2.5, 7.5, 12.5\}dB. We use the same hyperparameters as the ones used for fine-tuning the SER model with clean speech during adaptation. We tested two variations of our proposed text-guided environment-aware training: the CL-based representation \underline{TG-EAT-CL}, and the LLM-based representation \underline{TG-EAT-LLM}.

\subsection{Baselines}
\label{sec:baselines}

\noindent 
\underline{Original}: This model fine-tunes the model with clean emotional speech, with no adaptation to the noisy conditions.  

\noindent 
\underline{Retrain the original model with noisy speech (RT)}: This baseline updates the transformer encoder and the downstream head of the Original model with noisy speech. It does not use environmental information during adaptation and inference.  As described in Section \ref{ssec:data_preparation}, it uses 20 environmental conditions for adaptation. The evaluation uses 89 other environmental conditions.

\noindent 
\underline{Domain adversarial training (DAT)}: Inspired by Huang \etal \cite{Huang_2022}, we test a domain adversarial training strategy to adapt an SER model to multiple noisy conditions. Along with the downstream head for the SER task, we attach an environment classifier on top of the average-pooled transformer representations. The environment classifier has the same architecture as the downstream head for the SER task. The environment classifier is trained to minimize the cross-entropy loss between the predicted and the ground-truth noise types. We applied a \emph{gradient reversal layer} (GRL) between the environment classifier and the transformer encoder to train the transformer encoder to normalize the environment information in the resulting representations. Like the RT baseline, this baseline does not use environmental information during inference.

\noindent 
\underline{Enhance the noisy speech (SE)}: This baseline denoises the input noisy speech before feeding it into the original SER model. We use the \emph{frequency recurrent convolutional recurrent network} (FRCRN) framework \cite{Zhao_2022} to enhance the input speech. The FRCRN model is trained with the 4th DNS challenge dataset, achieving one of the top performances in this challenge \cite{Dubey_2022}.

\begin{table}[tb]
\centering
\caption{Average CCC for models using wav2vec2-large-robust feature vectors. We denote with {$^{\ast}$}, {$^{\dag}$}, {$^{\star}$}, and {$^{\ddag}$} when a model shows significantly better performance than the Original, RT, DAT, and SE models, respectively. We also mark {$^{\diamondsuit}$} and {$^{\clubsuit}$} when a baseline significantly perform better than the TG-EAT-CL and TG-EAT-LLM, respectively. We highlight in bold the best performance per condition.}
\label{tab:ccc-wav2vec2}
\begin{tabular*}{\linewidth}{@{\extracolsep{\fill}}ll|ccc}
\hline
\textbf{SNR} & \textbf{Model} & \textbf{Arousal} & \textbf{Dominance} & \textbf{Valence} \\
\hline
\hline
\multirow{6}{*}{\rotatebox[origin=c]{90}{\textbf{Clean}}}
 & Original (${\ast}$) & 0.63 & 0.53 & 0.41 \\
 & RT (${\dag}$) & 0.63 & 0.53 & 0.46$^{\ast}$ \\
 & DAT (${\star}$) & 0.63 & 0.51 & 0.45$^{\ast}$ \\
 & SE ({${\ddag}$}) & 0.53 & 0.48 & 0.37 \\
 & TG-EAT-CL ({${\diamondsuit}$}) & 0.63 & 0.53 & 0.45$^{\ast}$ \\
 & TG-EAT-LLM ({${\clubsuit}$}) & 0.63 & 0.53 & 0.46$^{\ast}$ \\
\hline
\multirow{6}{*}{\rotatebox[origin=c]{90}{\textbf{5dB}}}
 & Original (${\ast}$) & 0.60{$^{\ddag}$} & 0.51{$^{\ddag}$} & 0.40{$^{\ddag}$} \\
 & RT (${\dag}$) & \textbf{0.63}{$^{\ast}$}{$^{\ddag}$} & \textbf{0.52}{$^{\ddag}$} & 0.44{$^{\ast}$}{$^{\ddag}$} \\
 & DAT (${\star}$) & 0.62{$^{\ddag}$} & 0.50{$^{\ddag}$} & 0.44{$^{\ast}$}{$^{\ddag}$} \\
 & SE ({${\ddag}$}) & 0.50 & 0.44 & 0.35 \\
 & TG-EAT-CL ({${\diamondsuit}$}) & 0.62{$^{\ddag}$} & 0.51{$^{\ddag}$} & 0.45{$^{\ast}$}{$^{\ddag}$} \\
 & TG-EAT-LLM ({${\clubsuit}$}) & 0.62{$^{\ddag}$} & \textbf{0.52}{$^{\ddag}$} & \textbf{0.46}{$^{\ast}$}{$^{\ddag}$} \\
\hline
\multirow{6}{*}{\rotatebox[origin=c]{90}{\textbf{0dB}}}
 & Original (${\ast}$) & 0.54{$^{\ddag}$} & 0.46{$^{\ddag}$}{$^{\diamondsuit}$} & 0.31 \\
 & RT (${\dag}$) & 0.55{$^{\ddag}$}{$^{\diamondsuit}$} & 0.46{$^{\ddag}$}{$^{\diamondsuit}$} & 0.38{$^{\ast}$}{$^{\ddag}$} \\
 & DAT (${\star}$) & 0.54{$^{\ddag}$}{$^{\diamondsuit}$} & 0.44{$^{\ddag}$} & \textbf{0.39}{$^{\ast}$}{$^{\ddag}$} \\
 & SE ({${\ddag}$}) & 0.47 & 0.41 & 0.35{$^{\ast}$} \\
 & TG-EAT-CL ({${\diamondsuit}$}) & 0.52{$^{\ddag}$} & 0.42 & 0.38{$^{\ast}$}{$^{\ddag}$} \\
 & TG-EAT-LLM ({${\clubsuit}$}) & \textbf{0.56}{$^{\ast}$$^{\star}$}{$^{\ddag}$}{$^{\diamondsuit}$} & \textbf{0.47}{$^{\star}$}{$^{\ddag}$}{$^{\diamondsuit}$} & \textbf{0.39}{$^{\ast}$}{$^{\ddag}$} \\
\hline
\multirow{6}{*}{\rotatebox[origin=c]{90}{\textbf{-5dB}}}
 & Original (${\ast}$) & 0.26{$^{\diamondsuit}$} & 0.24{$^{\ddag}$}{$^{\diamondsuit}$} & 0.11 \\
 & RT (${\dag}$) & 0.22 & 0.21 & 0.15{$^{\ast}$$^{\star}$} \\
 & DAT (${\star}$) & 0.24{$^{\diamondsuit}$} & 0.22{$^{\ddag}$} & 0.13 \\
 & SE ({${\ddag}$}) & 0.23 & 0.19 & \textbf{0.19}{$^{\ast}$$^{\dag}$$^{\star}$}{$^{\diamondsuit}$}{$^{\clubsuit}$} \\
 & TG-EAT-CL ({${\diamondsuit}$}) & 0.21 & 0.20 & 0.15{$^{\ast}$$^{\star}$} \\
 & TG-EAT-LLM ({${\clubsuit}$}) & \textbf{0.28}{$^{\ast}$$^{\dag}$$^{\star}$}{$^{\diamondsuit}$} & \textbf{0.26}{$^{\ast}$$^{\dag}$$^{\star}$}{$^{\ddag}$}{$^{\diamondsuit}$} & 0.16{$^{\ast}$$^{\star}$} \\
\hline
\multirow{6}{*}{\rotatebox[origin=c]{90}{\textbf{Random}}}
 & Original (${\ast}$) & 0.47{$^{\ddag}$}{$^{\diamondsuit}$} & 0.41{$^{\ddag}$}{$^{\diamondsuit}$} & 0.29 \\
 & RT (${\dag}$) & 0.47{$^{\ddag}$}{$^{\diamondsuit}$} & 0.39{$^{\ddag}$}{$^{\diamondsuit}$} & 0.35$^{\ast}${$^{\ddag}$} \\
 & DAT (${\star}$) & 0.47{$^{\ddag}$}{$^{\diamondsuit}$} & 0.39{$^{\ddag}$} & 0.34$^{\ast}${$^{\ddag}$} \\
 & SE ({${\ddag}$}) & 0.37 & 0.32 & 0.30 \\
 & TG-EAT-CL ({${\diamondsuit}$}) & 0.44{$^{\ddag}$} & 0.37{$^{\ddag}$} & 0.34$^{\ast}${$^{\ddag}$} \\
 & TG-EAT-LLM ({${\clubsuit}$}) & \textbf{0.50}{$^{\ast}$$^{\dag}$$^{\star}$}{$^{\ddag}$}{$^{\diamondsuit}$} & \textbf{0.42}{$^{\dag}$$^{\star}$}{$^{\ddag}$}{$^{\diamondsuit}$} & \textbf{0.36}{$^{\ast}$$^{\star}$}{$^{\ddag}$} \\
\hline
\end{tabular*}
\end{table}

\begin{table}[tb]
\centering
\caption{Average CCC for models using wavlm-base-plus feature vectors. We use the same notations as in Table \ref{tab:ccc-wav2vec2}. We highlight in bold the best performance per condition.}
\label{tab:ccc-wavlm}
\begin{tabular*}{\linewidth}{@{\extracolsep{\fill}}ll|ccc}
\hline
\textbf{SNR} & \textbf{Model} & \textbf{Arousal} & \textbf{Dominance} & \textbf{Valence} \\
\hline
\hline
\multirow{6}{*}{\rotatebox[origin=c]{90}{\textbf{Clean}}}
 & Original (${\ast}$) & 0.60 & 0.49 & 0.46 \\
 & RT (${\dag}$) & 0.59 & 0.49 & 0.43 \\
 & DAT (${\star}$) & 0.58 & 0.48 & 0.48 \\
 & SE ({${\ddag}$}) & 0.58 & 0.46 & 0.43 \\
 & TG-EAT-CL ({${\diamondsuit}$}) & 0.57 & 0.47 & 0.47 \\
 & TG-EAT-LLM ({${\clubsuit}$}) & 0.59 & 0.48 & 0.46 \\
\hline
\multirow{6}{*}{\rotatebox[origin=c]{90}{\textbf{5dB}}}
 & Original (${\ast}$) & 0.54 & 0.45 & 0.44 \\
 & RT (${\dag}$) & \textbf{0.58$^{\ast}${$^{\ddag}$}} & \textbf{0.48$^{\ast}$}{$^{\ddag}$} & 0.41 \\
 & DAT (${\star}$) & \textbf{0.58$^{\ast}${$^{\ddag}$}} & \textbf{0.48$^{\ast}$}{$^{\ddag}$} & \textbf{0.47$^{\ast}$$^{\dag}$}{$^{\ddag}$} \\
 & SE ({${\ddag}$}) & 0.55 & 0.45 & 0.40 \\
 & TG-EAT-CL ({${\diamondsuit}$}) & 0.57$^{\ast}${$^{\ddag}$} & 0.47$^{\ast}${$^{\ddag}$} & 0.46$^{\ast}$$^{\dag}${$^{\ddag}$} \\
 & TG-EAT-LLM ({${\clubsuit}$}) & \textbf{0.58$^{\ast}$}{$^{\ddag}$} & 0.47$^{\ast}${$^{\ddag}$} & 0.44$^{\dag}${$^{\ddag}$} \\
\hline
\multirow{6}{*}{\rotatebox[origin=c]{90}{\textbf{0dB}}}
 & Original (${\ast}$) & 0.40 & 0.31 & 0.33 \\
 & RT (${\dag}$) & 0.53$^{\ast}$ & 0.43$^{\ast}$ & 0.33 \\
 & DAT (${\star}$) & 0.53$^{\ast}$ & \textbf{0.45}{$^{\ast}$$^{\dag}$}{$^{\diamondsuit}$} & \textbf{0.41}{$^{\ast}$$^{\dag}$} \\
 & SE ({${\ddag}$}) & 0.53$^{\ast}$ & 0.44$^{\ast}$ & \textbf{0.41}$^{\ast}$$^{\dag}$ \\
 & TG-EAT-CL ({${\diamondsuit}$}) & 0.51$^{\ast}$ & 0.42$^{\ast}$ & 0.40$^{\ast}$$^{\dag}$ \\
 & TG-EAT-LLM ({${\clubsuit}$}) & \textbf{0.55}{$^{\ast}$$^{\dag}$$^{\star}$}{$^{\ddag}$}{$^{\diamondsuit}$} & \textbf{0.45}$^{\ast}$$^{\dag}${$^{\diamondsuit}$} & 0.38$^{\ast}$$^{\dag}$ \\
\hline
\multirow{6}{*}{\rotatebox[origin=c]{90}{\textbf{-5dB}}}
 & Original (${\ast}$) & 0.11 & 0.07 & 0.10 \\
 & RT (${\dag}$) & 0.18{$^{\ast}$} & 0.11{$^{\ast}$} & 0.12 \\
 & DAT (${\star}$) & 0.22{$^{\ast}$$^{\dag}$}{$^{\diamondsuit}$} & 0.16{$^{\ast}$$^{\dag}$}{$^{\diamondsuit}$} & 0.17{$^{\ast}$$^{\dag}$} \\
 & SE ({${\ddag}$}) & 0.28{$^{\ast}$$^{\dag}$$^{\star}$}{$^{\diamondsuit}$} & \textbf{0.22}{$^{\ast}$$^{\dag}$$^{\star}$}{$^{\diamondsuit}$} & \textbf{0.23}{$^{\ast}$$^{\dag}$$^{\star}$}{$^{\diamondsuit}$}{$^{\clubsuit}$} \\
 & TG-EAT-CL ({${\diamondsuit}$}) & 0.17{$^{\ast}$} & 0.11{$^{\ast}$} & 0.18{$^{\ast}$$^{\dag}$} \\
 & TG-EAT-LLM ({${\clubsuit}$}) & \textbf{0.29}{$^{\ast}$$^{\dag}$$^{\star}$}{$^{\diamondsuit}$} & 0.20{$^{\ast}$$^{\dag}$$^{\star}$}{$^{\diamondsuit}$} & 0.20{$^{\ast}$$^{\dag}$$^{\star}$} \\
\hline
\multirow{6}{*}{\rotatebox[origin=c]{90}{\textbf{Random}}}
 & Original (${\ast}$) & 0.34 & 0.25 & 0.31 \\
 & RT (${\dag}$) & 0.45$^{\ast}$ & 0.33$^{\ast}$ & 0.30 \\
 & DAT (${\star}$) & 0.46{$^{\ast}$$^{\dag}$}{$^{\ddag}$}{$^{\diamondsuit}$} & 0.37{$^{\ast}$$^{\dag}$}{$^{\diamondsuit}$} & \textbf{0.38}{$^{\ast}$$^{\dag}$}{$^{\ddag}$} \\
 & SE ({${\ddag}$}) & 0.44$^{\ast}$ & 0.35$^{\ast}$ & 0.35$^{\ast}$ \\
 & TG-EAT-CL ({${\diamondsuit}$}) & 0.43$^{\ast}$ & 0.33$^{\ast}$ & 0.37$^{\ast}$$^{\dag}$ \\
 & TG-EAT-LLM ({${\clubsuit}$}) & \textbf{0.50}{$^{\ast}$$^{\dag}$$^{\star}$}{$^{\ddag}$}{$^{\diamondsuit}$} & \textbf{0.39}{$^{\ast}$$^{\dag}$$^{\star}$}{$^{\ddag}$}{$^{\diamondsuit}$} & 0.36{$^{\ast}$$^{\dag}$} \\
\hline
\end{tabular*}
\end{table}

\section{Results}
\label{sec:results}
\subsection{Emotion recognition performance}
\label{ssec:ser_performance}

We report the SER performance of our text-guided environment-aware training with our baselines. As described in Section~\ref{ssec:data_preparation}, we use ten different evaluation sets for three SNR levels. We report the average CCC of ten experiments for each SNR level. We conduct a one-tailed Welch's t-test between the baselines and our proposed models to assess if the training strategy shows significantly better SER performance in noisy conditions. We assert significance at $p$-value $<$ 0.05. 
 
Tables~\ref{tab:ccc-wav2vec2} and \ref{tab:ccc-wavlm} illustrate the SER performance of each model in noisy testing environments. When comparing our baselines (RT, DAT, SE) with the original model, they do not consistently yield performance improvement for all the attributes. RT does not improve performance for either arousal or dominance with the wav2vec2-large-robust feature vector, or for valence with the wavlm-base-plus feature vector. Although the DAT and SE show significant performance improvements with the wavlm-base-plus feature vector, both baselines fail to improve arousal and dominance prediction performance with the wav2vec2-large-robust feature vector. Since these baselines do not use environmental information, we can observe the importance of incorporating it when adapting the SER model to multiple noisy environments. 

Compared with the baselines, our proposed TG-EAT-LLM performs the best when using the wav2vec2-large-robust feature vector.
In the random condition, TG-EAT-LLM improves the original model's performance by 6.3\% (arousal), 2.4\% (dominance), and 24.1\% (valence). It yields the best performance with the wavlm-base-plus feature vector for arousal and dominance prediction tasks. In the random condition, TG-EAT-LLM shows performance gains of 8.6\% (arousal) and 5.4\% (dominance) compared to the best baseline, DAT. Unlike with the wav2vec2-large-robust representation, DAT significantly improves the original model's performance for all the attributes with the wavlm-base-plus representation. 
The wavlm-base-plus is pre-trained with noisy data, while the wav2vec2-large-robust is trained with a diverse speech corpus under clean conditions. This difference makes the wavlm-base-plus inherently more robust to noise, which leads to the successful improvement with the baselines that do not use environmental information.
We note that TG-EAT-LLM consistently outperforms the original model across all SSL representations. These results indicate that guiding the SER model with LLM-based representation can improve the noise-robustness for the SER task. It shows good generalization to unknown environments. 
 
For the valence prediction, the SE baseline shows the best performance under the -5dB condition. 
Previous studies have shown that valence performance correlates with the linguistic information \cite{Wagner_2023}. This phenomenon could explain how the SE baseline can improve valence performance by explicitly enhancing speech intelligibility. However, it does not always yield the best performance for arousal and dominance. Both arousal and dominance are related to acoustic characteristics rather than linguistic information. 
Thus, this observation implies that the enhancement module can manipulate the acoustic characteristics of the original speech. We can also see that the SE baseline does not yield the best performance for valence under 5dB conditions, where the impact of acoustic distortion could be higher than the impact of intelligibility improvement.

When we compare the TG-EAT-CL and TG-EAT-LLM models, we conclude that the CL-based representation does not show a performance improvement over the original SER model, especially with the wav2vec2-large-robust feature vector. We can clearly see that the TG-EAT-CL model does not improve the performance for arousal and dominance in the 0dB and -5dB conditions. This result indicates that pre-training the text encoder to have enriched semantic information is more helpful for the noise-robust SER model than pre-training the text encoder with an audio-text pair.

An interesting and counter-intuitive finding here is that the models trained with noisy speech (e.g., RT, DAT, TG-EAT-CL, TG-EAT-LLM) outperform the Original model under clean conditions when experimenting with the wav2vec2-large-robust architecture. We assume this improvement is caused by exposing the noisy speech to the wav2vec2-large-robust representation, which is not trained with noisy speech in its pre-training stage. Previous studies have shown that augmenting the training set with multiple conditions not only improves \emph{automatic speech recognition} (ASR) performance under noisy conditions but also improves under a clean condition \cite{Ko_2017, Pesoparada_2022}. As discussed in the previous section, we hypothesize that improvements in speech intelligibility leads to improvement of valence prediction. Based on these observations, we conclude that this phenomenon, while unintuitive, demonstrates the benefits of data augmentation under clean conditions.

\begin{figure}[tb]
\centering
{\includegraphics[width=\linewidth]{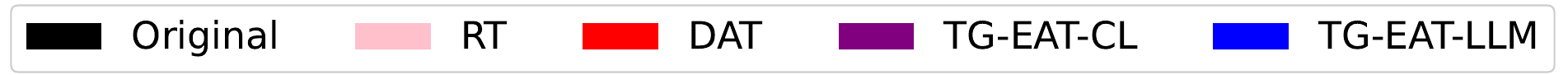}}
\begin{minipage}[b]{0.45\linewidth}

  \centering
  \centerline{\includegraphics[width=\linewidth]{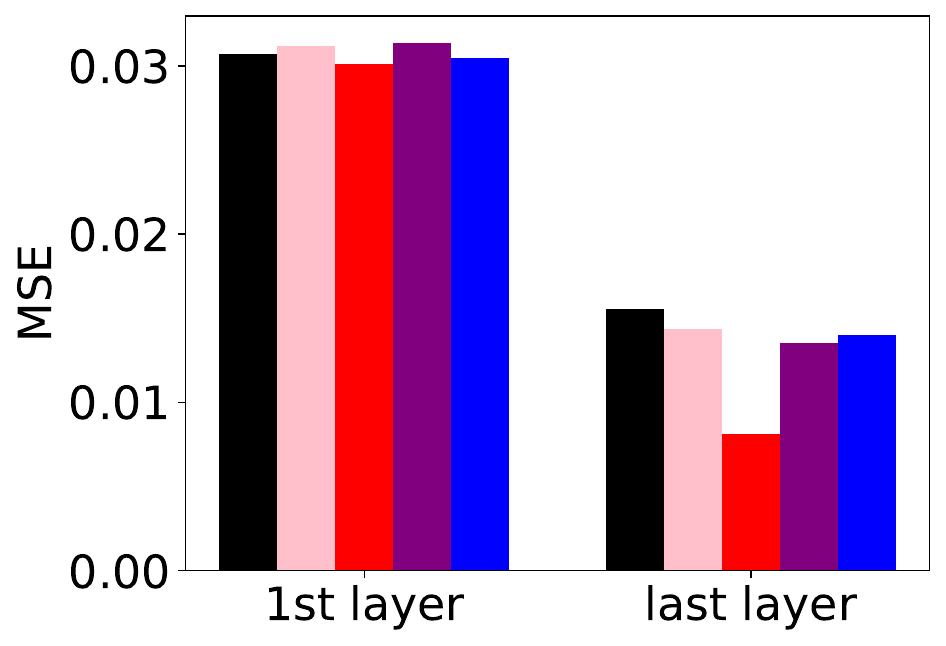}}
  \centerline{(a)}\medskip
\end{minipage}
\hfill
\begin{minipage}[b]{0.45\linewidth}
  \centering
  \centerline{\includegraphics[width=0.95\linewidth]{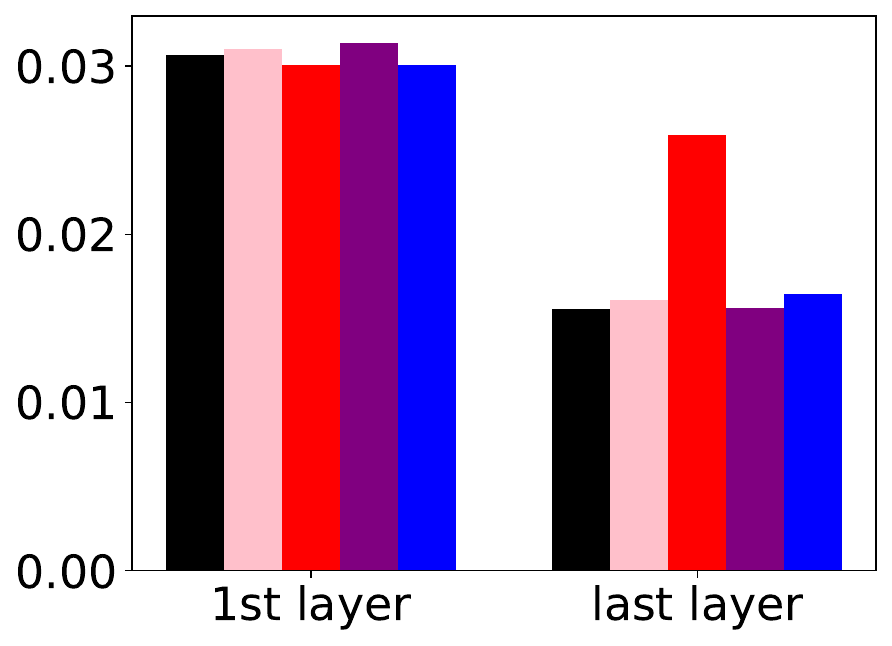}}
  \centerline{(b)}\medskip
\end{minipage}
\caption{ 
Embedding differences in the first and the last transformer encoder layers using clean and noisy speech in the -5dB condition. We use the wavlm-base-plus feature vector in this analysis. (a) illustrates the \emph{mean square error} (MSE) between the clean and noisy representations, where both representations are extracted from each of the final models. (b) illustrates the MSE between the clean representation extracted from the Original model and the noisy representation extracted from each final model.}

\label{fig:embedding_analysis}
\end{figure}

\subsection{Embedding analysis}
\label{ssec:embedding_analysis}
Section~\ref{ssec:ser_performance} demonstrated that the TG-EAT-LLM approach shows better performance than the environment-agnostic baselines and the TG-EAT-CL approach. Our initial assumption is that the proposed TG-EAT-LLM can learn appropriate denoising functions for the transformer encoder. To verify this assumption, we analyze the difference between the clean and noisy representations (Fig. \ref{fig:embedding_analysis}(a)). We use the wavlm-base-plus feature vector and the noisy speech from the -5dB condition for this analysis. The first analysis compares the clean and noisy representation extracted from each model. We want to assess with this analysis if the model is robust by comparing the representation obtained with clean and noisy speech. The second analysis compares the clean representation from the Original framework and the noisy representation from each of the models (Fig. \ref{fig:embedding_analysis}(b)). In this analysis, we want to assess if the model can keep the knowledge of the original SER model. We extract the representations from the first and the last transformer encoder layers and then calculate the mean square difference between clean and noisy representations for each layer.

Figure~\ref{fig:embedding_analysis} illustrates our analysis results. When extracting clean and noisy representations from the same model, we can first see that DAT shows the lowest difference in the last transformer layer. On the contrary, it shows the highest difference when extracting the clean representation from the original model. This result demonstrates the risk of catastrophic forgetting when using the DAT method. Although it can normalize the environmental difference in the adapted model, its representation can deviate from the original SER model's representation. However, our TG-EAT method does not highly increase the difference compared to the original model's clean representation. This result indicates that TG-EAT can minimize the risk of catastrophic forgetting during adaptation by introducing environmental information about the speech. 

Compared with the TG-EAT-LLM method, TG-EAT-CL shows a higher representation difference in the first layer. When comparing the clean and noisy representations from the same model, TG-EAT-LLM shows 7.7\% less representation difference than the TG-EAT-CL method in the first transformer layer. However, TG-EAT-CL shows less representation difference than the TG-EAT-LLM in the last layer. Even though the downstream head uses the representation from the last transformer layer, TG-EAT-CL shows worse performance than the TG-EAT-LLM approach. LLM-based representation can better denoise the acoustic representation than the CL-based representation. In addition, we speculate that the embedding difference in the lower transformer layer might be the crucial factor for increasing the robustness to noise of the SER system.

\begin{figure}[tb]
	\centering
        \includegraphics[width=0.6\linewidth]{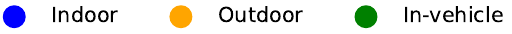}
	\subfloat[TG-EAT-CL]{
		\includegraphics[width=0.45\linewidth]{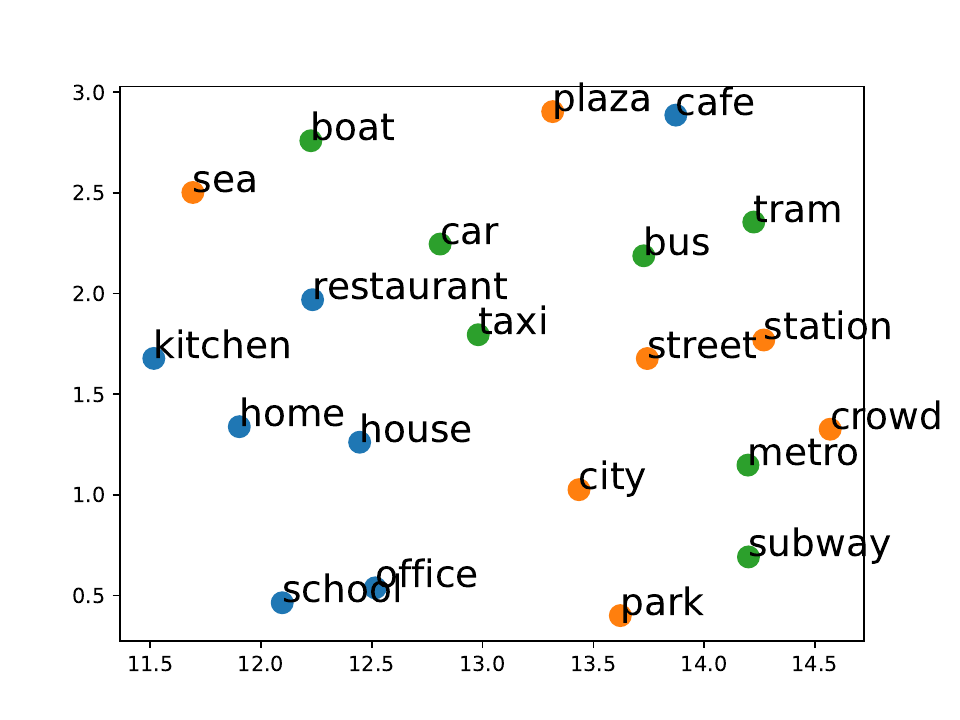}
	}
	\subfloat[TG-EAT-LLM]{
		\includegraphics[width=0.45\linewidth]{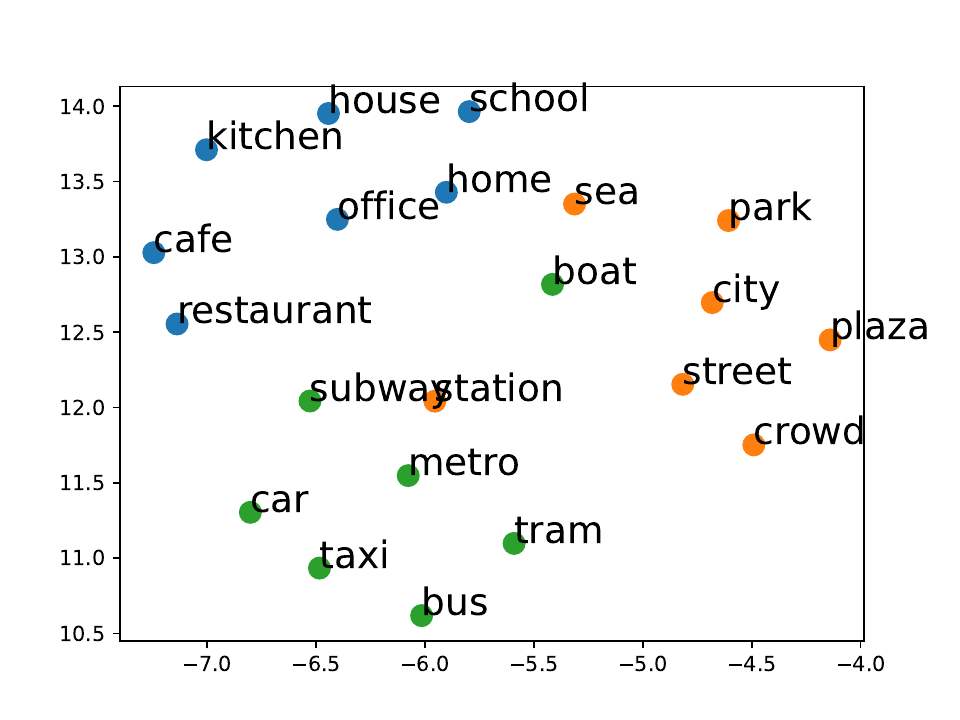}
	}
	\caption{Visualization of text-based environment embeddings. We use UMAP to project text embeddings into 2D space.}
	\label{fig:umap_analysis}
\end{figure}

We also investigate if the proposed text-based environment embedding clusters similar environments together, which is the key premise of the proposed approach to deal with unseen environments. First, we randomly select 21 different keywords, each representing an indoor, outdoor, and in-vehicle environment. Each environment includes seven keywords extracted from the train and test sets, aiming to illustrate the model's capability to cluster similar environments in both seen and unseen environments. We extract the text embedding from these 21 keywords by using the same template that we used for our TG-EAT frameworks (i.e., \emph{``This speech is recorded in \underline{\{environment\}}.''})
We project these embeddings into the 2D space to visualize the embedding space using the \emph{uniform manifold approximation and projection} (UMAP) method \cite{Mcinnes_2018}. Figure \ref{fig:umap_analysis} illustrates the text embedding space of TG-EAT-CL and TG-EAT-LLM. The figure shows that both frameworks cluster semantically similar environmental conditions together. For example, we observe the embeddings for ``boat'' and ``sea,'' together. We also observe the ones for ``subway'' and ``station'' clustered together. Both encoders cluster the house environments (``house'', ``home'', ``kitchen'') and the vehicle environments (``bus'', ``taxi'', ``car''), which indicates that the text encoder can cluster acoustically similar environments. This analysis implies that our proposed frameworks can handle unseen environments by clustering acoustically and semantically similar environments.

\begin{table}[tb] 
\caption{Silhouette score of text embedding space of TG-EAT-CL and TG-EAT-LLM. We apply K-means clustering on the projected environmental embedding with $K$ clusters ($K={3, 5, 7}$).}
\label{tab:txt_silhouette_score} 
\centering 
\begin{tabular*}{\columnwidth}{@{\extracolsep{\fill}}l|ccc} 
\hline 
& K = 3 & K = 5 & K = 7  \\  
\hline 
 \hline 
TG-EAT-CL & 0.11 & 0.10 & 0.10 \\
TG-EAT-LLM & 0.57 & 0.43 & 0.40  \\
\hline 
\end{tabular*} 
\end{table}

To provide a quantitative analysis of the environmental embeddings and their impact on the model's representations, we evaluated the clustering quality of the environmental text embeddings. We extracted embeddings for all environments listed in Table \ref{tab:keywords} from both the TG-EAT-CL and TG-EAT-LLM text encoders. We calculated the silhouette score for each set of embeddings using the K-means clustering \cite{Shahapure_2020}. Table \ref{tab:txt_silhouette_score} illustrates the silhouette score of each embedding projection with a different number of clusters. With three clusters, the TG-EAT-LLM embeddings achieved a score of 0.57, substantially higher than the 0.10 score from the TG-EAT-CL embeddings. This result indicates that the LLM-based encoder generates more separable and well-defined clusters for different environments. This higher-quality embedding structure correlates with the superior performance of the TG-EAT-LLM model in noisy conditions, suggesting that more discriminable environmental representations are key to achieving robust performance.

\begin{table}[tb] 
\caption{Average CCC of the ten experiments for the seen environment. The environmental conditions for the train set and the test set are the same. We compare the proposed method with the baselines by using the wavlm-base-plus model.} 
\label{tab:seen_environment} 

\centering 
\begin{tabular*}{0.95\columnwidth}{@{\extracolsep{\fill}}ll|ccc} 
\hline 
SNR & Model & Arousal & Dominance & Valence \\  
\hline 
\hline 
\centering{\multirow{3}{*}{\rotatebox[origin=c]{90}{5dB}}} 
 & Original & 0.54 & 0.46 & 0.45 \\ 
 & One-hot & 0.59 & 0.48 & 0.47 \\ 
 & TG-EAT-LLM & 0.59 & 0.48 & 0.47 \\ 
 \hline 
\centering{\multirow{3}{*}{\rotatebox[origin=c]{90}{0dB}}} 
 & Original & 0.40 & 0.32 & 0.35 \\ 
 & One-hot & 0.56 & 0.45 & 0.42 \\ 
 & TG-EAT-LLM & 0.56 & 0.46 & 0.40 \\ 
 \hline 
\centering{\multirow{3}{*}{\rotatebox[origin=c]{90}{-5dB}}} 
 & Original & 0.09 & 0.06 & 0.10 \\ 
 & One-hot & 0.29 & 0.20 & 0.21 \\ 
 & TG-EAT-LLM & 0.27 & 0.18 & 0.21 \\ 
 \hline 
\end{tabular*} 
\end{table}

\begin{table}[tb] 
\caption{Average CCC of the ten experiments for the unseen environment. We compare the proposed method with the baselines by using the wavlm-base-plus model. We denote with {$^{\ast}$} when a model shows significantly better performance than the Original model.} 
\label{tab:unseen_environment} 
\centering 
\begin{tabular*}{\columnwidth}{@{\extracolsep{\fill}}ll|ccc} 
\hline 
SNR & Model & Arousal & Dominance & Valence \\ 
\hline 
\hline 
\centering{\multirow{6}{*}{\rotatebox[origin=c]{90}{5dB}}} 
 & Original & 0.54 & 0.45 & \textbf{0.44} \\ 
 & RT & 0.58$^{\ast}$ & 0.48$^{\ast}$ & 0.41 \\
 & GloVe & 0.58$^{\ast}$ & 0.47$^{\ast}$& 0.42 \\ 
 & AST & \textbf{0.59}$^{\ast}$ & \textbf{0.49}$^{\ast}$ & 0.41 \\
 & TG-EAT-LLM & 0.58$^{\ast}$ & 0.48$^{\ast}$ & \textbf{0.44} \\
 \hline 
\centering{\multirow{6}{*}{\rotatebox[origin=c]{90}{0dB}}} 
 & Original & 0.40 & 0.31 & 0.33 \\ 
 & RT & 0.53$^{\ast}$ & 0.43$^{\ast}$ & 0.33 \\
 & GloVe & 0.53$^{\ast}$ & 0.42$^{\ast}$ & 0.37$^{\ast}$ \\ 
 & AST & \textbf{0.55}$^{\ast}$ & 0.44$^{\ast}$ & 0.34 \\
 & TG-EAT-LLM & \textbf{0.55}$^{\ast}$ & \textbf{0.45}$^{\ast}$ & \textbf{0.38}$^{\ast}$ \\
 \hline 
\centering{\multirow{6}{*}{\rotatebox[origin=c]{90}{-5dB}}} 
& Original & 0.11 & 0.07 & 0.10 \\ 
 & RT & 0.18$^{\ast}$ & 0.11$^{\ast}$ & 0.12 \\
 & GloVe & 0.24$^{\ast}$ & 0.16$^{\ast}$ & 0.18$^{\ast}$ \\ 
 & AST & 0.28$^{\ast}$ & \textbf{0.20}$^{\ast}$ & 0.14$^{\ast}$ \\
 & TG-EAT-LLM & \textbf{0.29}$^{\ast}$ & \textbf{0.20}$^{\ast}$ & \textbf{0.20}$^{\ast}$ \\
 \hline 
\end{tabular*} 
\end{table}

\subsection{Evaluation of Different Types of Environmental Embedding}
\label{ssec:env_emb}

Our proposed method uses the embedding extracted from the text encoder to represent the testing environmental condition. To verify the benefits of using a text-based environmental embedding, we compare it with three different types of environmental embedding: \emph{one-hot encoding} (One-hot), \emph{global vectors for word representation} (GloVe) \cite{Pennington_2014}, and \emph{audio spectrogram transformer representation} (AST)  \cite{Gong_2021}. One-hot uses 20-dimensional binary vectors, where 1 represents the target environment condition, and 0 represents the others. Each dimension corresponds to the environmental condition of the training set. This embedding fully represents a seen environment with a simple vector; however, it cannot represent unseen environments, which is inappropriate for real-world services. GloVe is a word-level vector representation extracted from the regression model that considers the co-occurrences of words. We import the pre-trained GloVe vector collections, which consist of a 2.2 million-word vocabulary. We select the word vector representation that corresponds to the target noisy environment. The resulting representation is a 300-dimensional vector. This representation can handle unseen environments through text description, but it is semantically limited compared to our proposed text encoders. AST uses a transformer architecture to map the spectrogram patches into an audio-level representation. The model is fine-tuned with sound event classification tasks using AudioSet, which serves as the noise sound corpus for our training set. We directly import the pre-trained checkpoint from HuggingFace and extract the patch-wise embedding sequence from the given input. We apply average pooling to the extracted sequence to yield a single environment embedding, which is then fused with the pre-trained SER model. We do not fine-tune the pre-trained checkpoint jointly with the SER model, following the same strategy we use to train the TG-EAT-LLM framework. This model can automatically capture the acoustic characteristics from the audio-only input. However, it cannot explicitly use the semantic information of the testing environment.

We compare our proposed method with the one-hot vector in the seen environment scenario (Table \ref{tab:seen_environment}) and with the other baselines in the unseen environment scenario (Table \ref{tab:unseen_environment}). For the seen environment scenario, we used the same environmental conditions as the train set to contaminate the clean test set, but with different audio samples. We use ten different test sets and report the average CCC for both cases. Tables \ref{tab:seen_environment} and \ref{tab:unseen_environment} report the results for the seen and unseen environments, respectively. In the seen environment, our proposed method and the one-hot environment encoding model improve the original SER performance for all the conditions and attributes. Both models show similar performances in the seen environments. However, the one-hot encoding cannot cover unseen environments. This result demonstrates that the proposed text embedding can deal with both seen and unseen environments. Compared to the model that uses GloVe embeddings, our proposed method shows better SER performances in the 0dB and -5dB conditions. It also shows a better performance for valence in the 5dB condition. The GloVe model only considers word co-occurrence to get a word embedding, while our proposed text encoder model is pre-trained to understand the semantic information of a sentence. This result implies the importance of pre-training the text encoder with language modeling to get a robust environment embedding for performance improvement. The AST strategy significantly improves the performance for arousal and dominance. However, it fails to improve the performance for valence when the SNR level is high (e.g., 5dB and 0dB conditions). AST does not use semantic information from the testing environment to get environmental embedding; instead, it extracts the environmental information from the given audio. We hypothesize that AST confuses the environmental condition when the background noise amplitude is comparably lower than the speech sound. Unlike this approach, our proposed model relies on the text description, which is independent of the SNR level. Therefore, it performs better than AST for valence. In the 0 dB and {-5}~dB conditions, our method significantly improves the original models' performances for all the attributes. Considering that those low SNR levels are not presented while training the model, the result demonstrates that our proposed method is robust against unseen SNR levels, which is practical for real-world scenarios.

\begin{figure}[tb]
	\centering
        \includegraphics[width=0.6\linewidth]{figure/legend_txt.pdf}
	\subfloat[TG-EAT-CL]{
		\includegraphics[width=0.45\linewidth]{figure/txt_check_clap.pdf}
	}
	\subfloat[TG-EAT-CL-FT]{
		\includegraphics[width=0.45\linewidth]{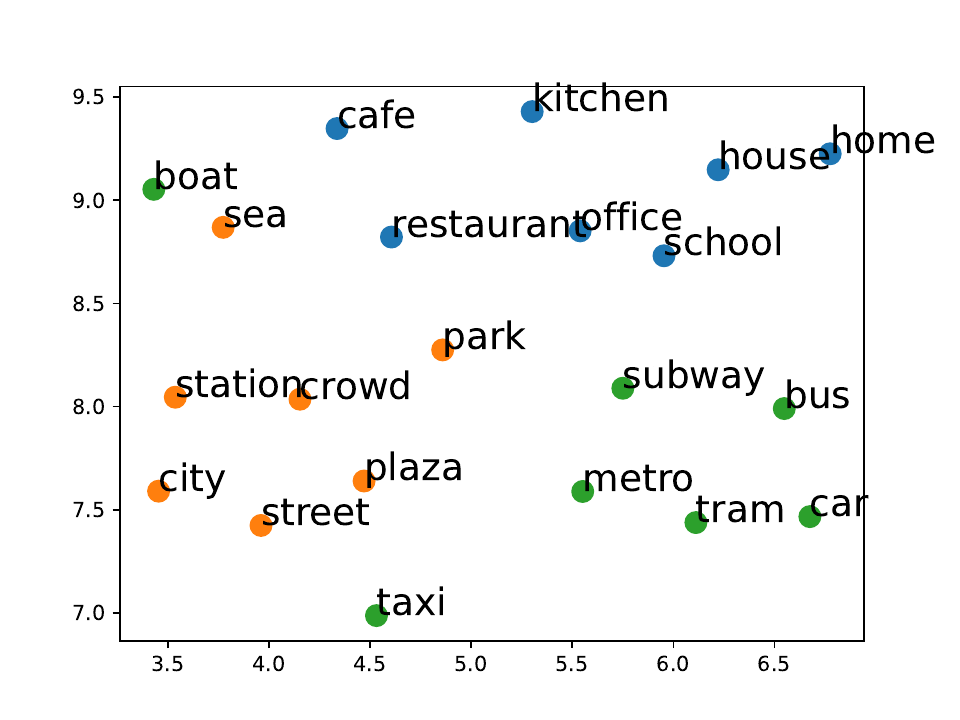}
	}
	\caption{Comparison of the text embedding projection obtained before fine-tuning (TG-EAT-CL) and after fine-tuning (TG-EAT-CL-FT). Similar to the plots in Figure \ref{fig:umap_analysis}, we use UMAP to project text embeddings into a 2D space.}
	\label{fig:umap_analysis_cl}
\end{figure}

\begin{figure*}[t]
\begin{minipage}[b]{0.3\linewidth}
  \centering
  \centerline{\includegraphics[width=6.0cm]{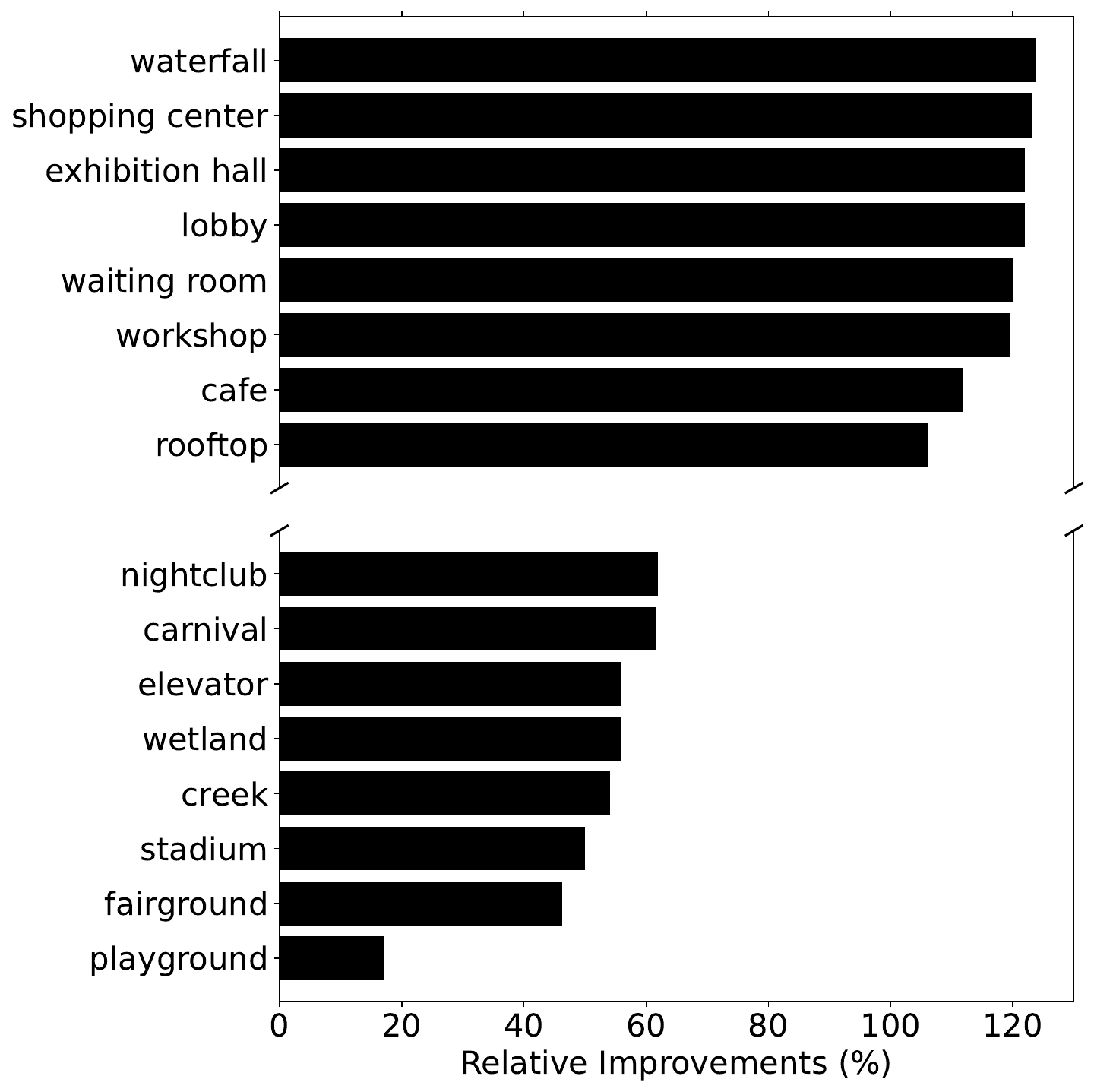}}
  \centerline{(a) Arousal}\medskip
\end{minipage}
\hfill
\begin{minipage}[b]{0.3\linewidth}
  \centering
  \centerline{\includegraphics[width=6.0cm]{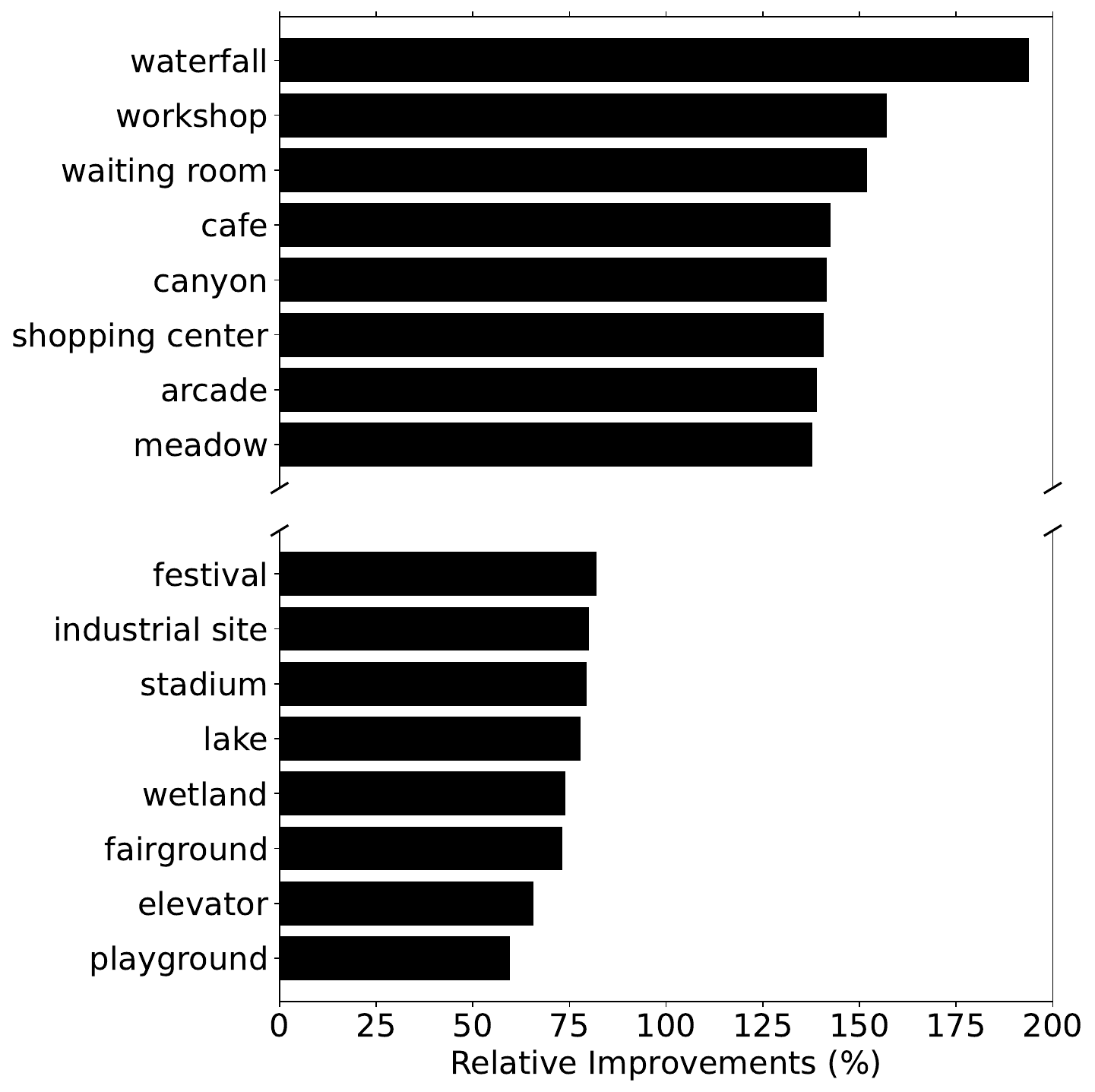}}
  \centerline{(b) Dominance}\medskip
\end{minipage}
\hfill
\begin{minipage}[b]{0.3\linewidth}
  \centering
 \centerline{\includegraphics[width=6.0cm]{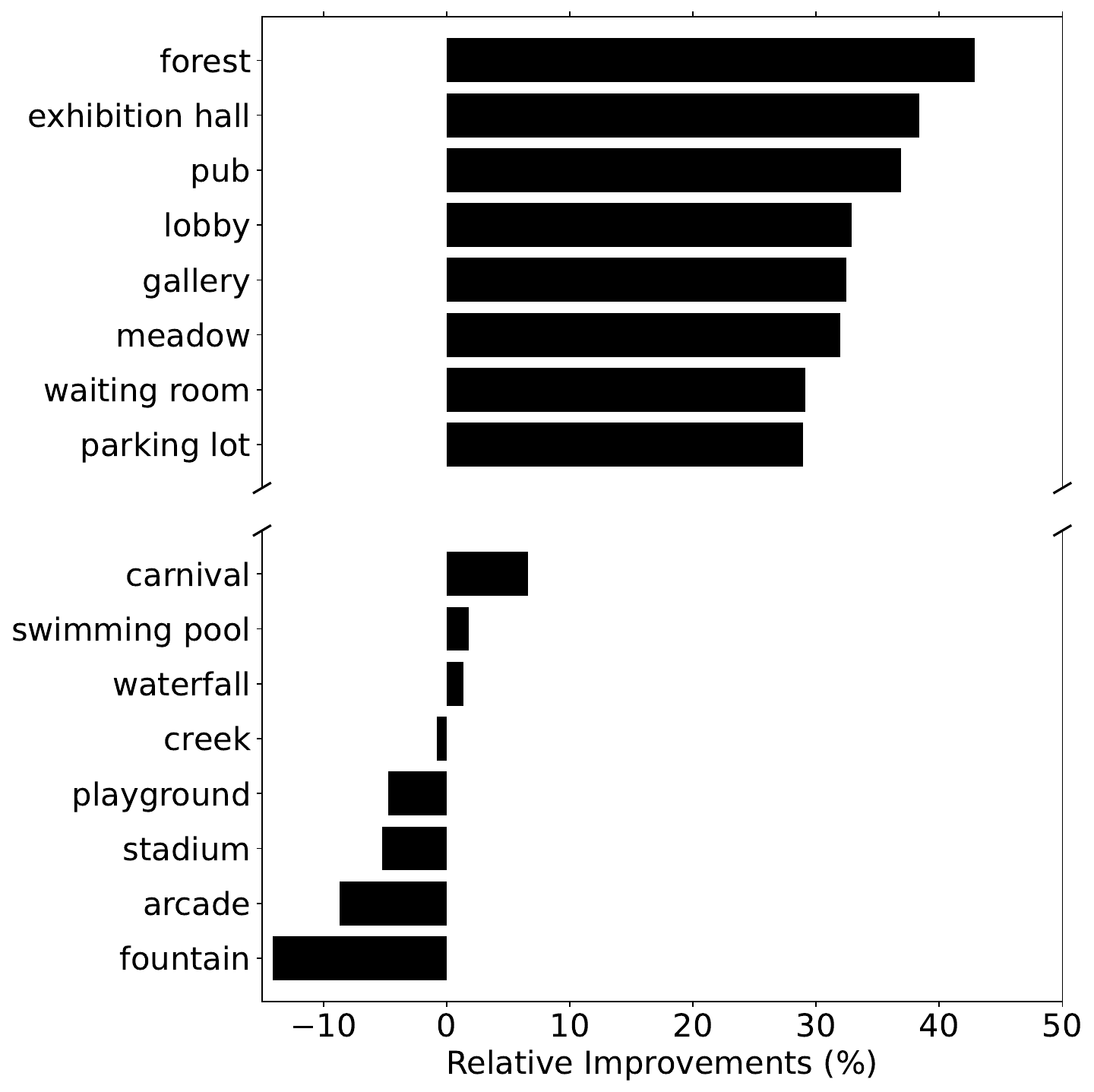}}
  \centerline{(c) Valence} \medskip
\end{minipage}

\caption{Relative improvement of fine-tuning the text encoder (TG-EAT-CL-FT) in the TG-EAT-CL framework under the -5dB condition. We illustrate 16 environments, including the eight highest and the eight lowest improvements for each attribute.}
\label{fig:imp_each_env}
\end{figure*}

\subsection{Benefit of Fine-Tuning the Text Encoder}
\label{ssec:text_encoder_fine_tuning}

Our results demonstrate that using the text encoder pre-trained with the CLAP strategy shows worse SER performance than using the pre-trained LLM. Despite this observation, we assume that this type of text encoder should have the potential to improve since the text encoder is pre-trained with the audio modality. Our assumption is that jointly fine-tuning the text encoder with the SER model could further improve the performance. Therefore, we compare the performance of an SER model by either freezing the text encoder or updating the encoder while adapting the SER model with the text-based environment embedding. We refer to the models that fine-tune the text encoder of the \emph{TG-EAT-CL} and \emph{TG-EAT-LLM} approaches during adaptation as \emph{TG-EAT-CL-FT} and \emph{TG-EAT-LLM-FT}, respectively. 

Table \ref{tab:fine_tuning} reports the average CCC of ten different test sets for each model. When comparing the \emph{TG-EAT-LLM} and \emph{TG-EAT-LLM-FT} implementations, they do not show significantly different performance. However, the \emph{TG-EAT-CL-FT} approach shows meaningful performance improvement over the \emph{TG-EAT-CL} implementation. For the -5dB conditions, it even reaches the best performance among all the models. When compared with TG-EAT-CL, fine-tuning the text encoder improves the recognition performance by 76.4\% (arousal), 100.0\% (dominance), and 27.7\% (valence). To analyze how fine-tuning benefits the model, we visualize in Figure \ref{fig:umap_analysis_cl} the text embedding projection from the text encoder used for the TG-EAT-CL and TG-EAT-CL-FT models. We use the same experiment setting as we used for illustrating Figure \ref{fig:umap_analysis}. We can see that some of the embeddings that were not clustered well in TG-EAT-CL are corrected in TG-EAT-CL-FT. For example, ``cafe'' is distant from other indoor environments in TG-EAT-CL. However, when fine-tuning the text encoder, its embedding gets closer to those environments. We also observe that such cluster alignments could lead to performance improvement. Figure \ref{fig:imp_each_env} illustrates the relative performance improvement of TG-EAT-CL-FT for each environmental condition. We can see that the TG-EAT-CL-FT model improves performance in the ``cafe'' environments. We can also see that the fine-tuning strategy can improve the performance for all the attributes, except for five environments in valence (``creek'', ``playground'', ``stadium'', ``arcade'', and ``fountain''). This observation illustrates the importance of compensating for the gap in the embedding space between the pre-trained text encoder space and the acoustic embedding. Although jointly fine-tuning the text encoder and the SER model can cost more memory space and computation time for the adaptation, this strategy can fully utilize the potential of the text encoder pre-trained with the audio modality.

\begin{table}[tb] 
\caption{Comparison of freezing the text encoder and updating it while adapting the SER model for the TG-EAT-CL and the TG-EAT-LLM models. We report the average CCC of the ten experiments for all the methods. We implement all the approaches with wavlm-base-plus feature vectors. We highlight in bold the best performance per condition.} 
\label{tab:fine_tuning} 
\centering 

\begin{tabular*}{0.95\columnwidth}{@{\extracolsep{\fill}}ll|ccc} 
\hline 
SNR & Model & Arousal & Dominance & Valence \\  
\hline 
\hline 
\centering{\multirow{4}{*}{\rotatebox[origin=c]{90}{5dB}}} 
 & TG-EAT-CL & 0.57 & 0.47 & 0.46 \\
 & TG-EAT-CL-FT & \textbf{0.58} & \textbf{0.48} & \textbf{0.48} \\
 & TG-EAT-LLM & \textbf{0.58} & 0.47 & 0.44 \\
 & TG-EAT-LLM-FT & 0.57 & 0.46 & 0.46 \\
 \hline 
\centering{\multirow{4}{*}{\rotatebox[origin=c]{90}{0dB}}} 
 & TG-EAT-CL & 0.51 & 0.42 & 0.40 \\
 & TG-EAT-CL-FT & \textbf{0.55} & \textbf{0.45} & \textbf{0.44} \\
 & TG-EAT-LLM & \textbf{0.55} & \textbf{0.45} & 0.38 \\
 & TG-EAT-LLM-FT & 0.54 & 0.44 & 0.41 \\
 \hline 
\centering{\multirow{4}{*}{\rotatebox[origin=c]{90}{-5dB}}} 
& TG-EAT-CL & 0.17 & 0.11 & 0.18 \\
 & TG-EAT-CL-FT & \textbf{0.30} & \textbf{0.22} & \textbf{0.23} \\
 & TG-EAT-LLM & 0.29 & 0.20 & 0.20 \\
 & TG-EAT-LLM-FT & 0.27 & 0.19 & 0.21 \\
 \hline 
\end{tabular*} 
\end{table}

\begin{table}[tb]
\centering
\caption{Average CCC of the six sessions with the clean and noisy version of the MSP-IMPROV corpus. We compared the proposed method with the baselines by using the wavlm-base-plus model.}
\label{tab:ccc-improv}
\begin{tabular*}{\linewidth}{@{\extracolsep{\fill}}ll|ccc}
\hline
\textbf{SNR} & \textbf{Model} & \textbf{Arousal} & \textbf{Dominance} & \textbf{Valence} \\
\hline
\hline
\multirow{4}{*}{\rotatebox[origin=c]{90}{\textbf{Clean}}}
 & Original & 0.38 & 0.44 & 0.41 \\
 & RT & 0.39 & 0.44 & 0.40 \\
 & TG-EAT-CL & 0.38 & 0.45 & \textbf{0.44} \\
 & TG-EAT-LLM & \textbf{0.40} & 0.45 & 0.42 \\
\hline
\multirow{4}{*}{\rotatebox[origin=c]{90}{\textbf{Random}}}
 & Original & 0.32 & 0.36 & 0.25 \\
 & RT & \textbf{0.40} & 0.42 & 0.30 \\
 & TG-EAT-CL & 0.36 & 0.42 & \textbf{0.34} \\
 & TG-EAT-LLM & \textbf{0.40} & 0.42 & 0.32 \\
\hline
\end{tabular*}
\end{table}

\subsection{Cross-corpus Generalization}
\label{ssec:cross_corpus}
To evaluate the generalization ability of our proposed TG-EAT models under unseen, out-of-domain dataset, we conduct a cross-corpus analysis. We test our baselines and the proposed TG-EAT models on the MSP-IMPROV dataset \cite{Busso_2017}, where their data acquisition process is different from our training set, the MSP-Podcast Corpus. For this evaluation, we evaluate the performance for each of the six sessions in the MSP-IMPROV corpus. We create a noisy version of this test set by contaminating the clean audio with noise from the DEMAND database at a randomly selected SNR level between -5dB and 5dB. 

Table \ref{tab:ccc-improv} illustrates our experiment results. Under the clean condition, we can see that our proposed TG-EAT models do not degrade the performance of baselines. Indeed, they outperform the RT baseline for valence. As we discussed in Section \ref{ssec:ser_performance}, training with multiple conditions could help performance improvement for valence even under the clean condition. We can see that guiding the model with environmental conditions can keep this benefit under a cross-corpus scenario, while preserving the Original SER model's generalization ability.

As expected, under noisy conditions, all models retrained with noisy speech (RT, TG-EAT-CL, and TG-EAT-LLM) significantly outperform the Original model, confirming that noise-aware training is crucial for robust performance on out-of-set corpora. Our models show a clear improvement in valence. As discussed in Section \ref{ssec:ser_performance}, arousal and dominance are more closely related to acoustic characteristics, while valence is correlated with linguistic content. By providing explicit environmental information, our TG-EAT framework may allow the SER model to better normalize for acoustic variability, thereby improving the extraction of linguistic content critical for valence prediction. 

Interestingly, the TG-EAT-CL model yields the best valence performance under noisy, out-of-set conditions, despite not being the top performer in the in-set evaluation. This result indicates that its learned representation possesses a strong capability for generalization, particularly under challenging mismatched conditions. While our framework shows clear benefits, we note that the significant improvements observed for arousal and dominance in the in-set condition did not fully transfer to this cross-corpus task. This suggests that future work could explore methods to further enhance the generalization of acoustically-related emotion attributes in out-of-domain scenarios.

\subsection{Impact of Mislabeled Audio Tags}
\label{ssec:miss_labeled_audio}
In a real-world scenario, location tags could be mislabeled due to an inaccurate GPS signal or ambiguous locations. To evaluate the impact of mislabeling audio tags, we present an ablation study to report the performance of the proposed TG-EAT-LLM framework, manipulating the environmental description. To simulate mislabeling, we intentionally manipulated the environmental tags associated with the input noisy speech at varying levels of distortion. Specifically, the environmental tags were randomly replaced with ratios of 25\%, 50\%, and 75\%. The original model trained without any mislabeled tags (0\% manipulation) was used as the baseline for comparison.

\begin{figure}
    \centering
    \includegraphics[width=0.7\linewidth]{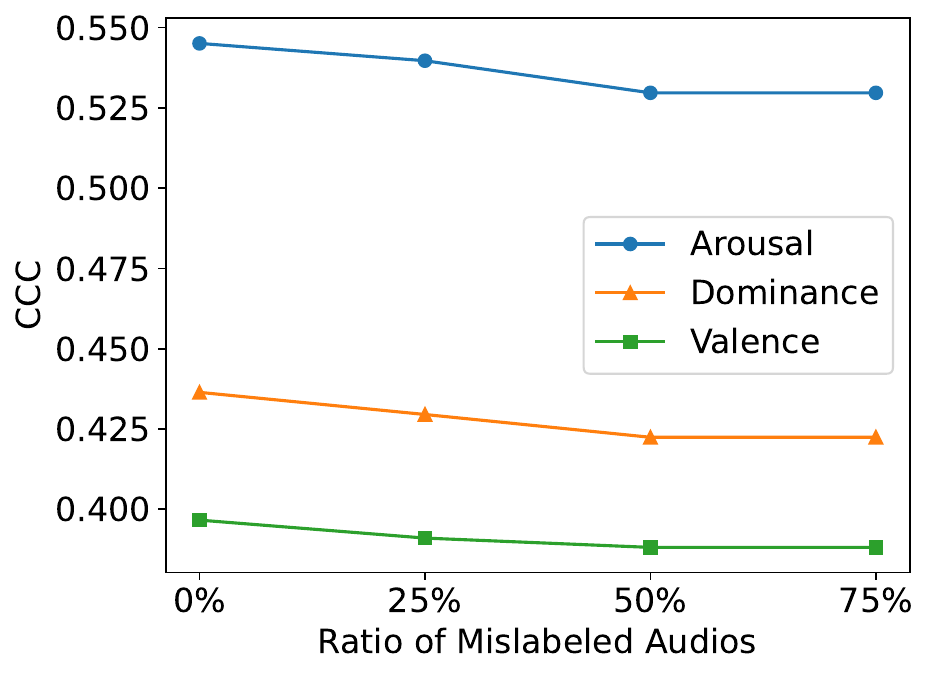}
    \caption{CCC of the TG-EAT-LLM model trained with noisy speech paired with mislabeled environment conditions.}
    \label{fig:miss_ratio}
\end{figure}

To ensure a robust evaluation, we tested the model on 10 different testing sets, each contaminated with noise at a randomly selected SNR between 5dB, 0dB, and -5 dB. The environmental conditions in the testing data were also chosen randomly to simulate diverse real-world scenarios. The average CCC across these 10 sets was computed for each manipulation ratio. Figure \ref{fig:miss_ratio} illustrates the result of the ablation study. The performance of the TG-EAT-LLM model gradually decreases as the levels of mislabeled audio tags increase. The trend indicates that as the model is exposed to higher degrees of mislabeling, it struggles to cluster recordings accurately from similar environmental conditions. This performance degradation highlights the sensitivity of the TG-EAT-LLM framework to the quality of audio tags and the need for accurate labeling during training. We can see that environmental conditions play a significant role in our proposed framework. The model relies heavily on this information to achieve robust SER performance, highlighting the importance of minimizing labeling errors when describing environmental conditions.

\begin{table}[tb] 
\caption{Silhouette score of the last hidden layer's embedding space of the RT and TG-EAT-LLM models.}
\label{tab:txt_conditions} 
\centering 
\begin{tabular*}{\columnwidth}{@{\extracolsep{\fill}}l|cc} 
\hline 
& RT & TG-EAT-LLM  \\  
\hline 
 \hline 
 Using the correct environmental tag & 0.363 & 0.133  \\
 Using a semantically similar but incorrect tag & 0.363 & 0.133  \\
 Using a semantically different tag & 0.360 & 0.140  \\
 Using without an environmental tag & 0.361 & 0.138  \\
\hline 
\end{tabular*} 
\end{table}

We investigate the degree of impact on our proposed model of using an audio tag that is mislabeled but either semantically similar or completely dissimilar to the correct environmental information. We measured the separability between the final layer embeddings of speech under clean versus noisy conditions. A robust model should normalize the environmental difference, which should lead to a low clustering quality when the embedding space is clustered by environmental differences. We compared our TG-EAT-LLM model against the RT baseline under four prompt conditions: (1) the correct environmental tag, (2) a semantically similar but incorrect tag, (3) a semantically different tag, and (4) no tag at all. We select five similar conditions (``kitchen'', ``house'', ``living room'', ``school'', ``office''), contaminating the test set with these noise types. SNR levels are randomly chosen from -5dB to 5dB. Condition (1) uses the same tag as the noise label in the input audio. Condition (2) randomly selects the tags from the four other similar conditions. Condition (3) randomly selects tags from five different tags that are semantically different from this group (``playground'', ``subway station'', ``town square'', ``construction site'', ``sports field''). Condition (4) does not use any environmental tags (i.e., the model only accepts the input audio).

Table \ref{tab:txt_conditions} shows the result of our experiment. The TG-EAT-LLM model consistently achieves much lower silhouette scores than the RT baseline. When using a correct tag, the RT baseline yields a 0.363 score, while our TG-EAT-LLM's score achieves a 0.133 score. This result illustrates that our framework effectively normalizes environmental differences from the speech representation. When compared to using the correct tag, using a semantically similar tag does not change the clustering score in TG-EAT-LLM. The score slightly increases when a semantically different tag or no tag is provided. The RT baseline's score remains unchanged regardless of the text input. These observations imply that guiding the model with environmental information during training can introduce sensitivity to semantically incorrect environmental information. The model does not ignore the prompt but instead uses it as intended to disentangle environmental noise from emotional content.

\subsection{Limitations}
\label{ssec:limitations}
Our proposed TG-EAT framework heavily relies on the assumption that the recorded speech is paired with accurate GPS location data, which is crucial for acquiring accurate environmental tags. However, the recorded speech could be associated with inaccurate or missing GPS points in real-world scenarios, leading to irrelevant or unavailable environmental tags. As discussed in Section \ref{ssec:miss_labeled_audio}, having irrelevant tags could degrade our system's performance, and missing tags would not provide the information for our model to work properly. Additionally, GIS mashups may fail to retrieve meaningful tags in areas with sparse or incomplete annotations, further limiting the system's ability to leverage the information of the recording conditions.  
Furthermore, even with accurate data, a single static tag may be an oversimplification for complex acoustic scenes with overlapping speech or rapidly changing soundscapes, potentially degrading performance. Those limitations demonstrate the need for future work to address potential inaccuracies in GPS data and missing GPS modality. In addition to the availability of accurate GPS information, our architectural approach to fuse the text embedding to the SER model was limited to concatenating a text embedding to audio tokens. The exploration of more dynamic fusion strategies \cite{Perez_2018, Alayrac_2022} remains a key area for future work to potentially build upon our findings and further enhance performance.

\section{Conclusions}
\label{sec:conclusion}

We proposed the TG-EAT method, which uses a text description of the testing environment for noise-robust SER.  
This approach inserts a text-based environment representation into an SER model, leading it to improve the prediction with respect to the given environmental information. Our experiment demonstrated that the LLM-based representation can improve SER performance under noisy conditions, especially when dealing with low SNR conditions. Our analysis indicates that the pre-trained text encoder can cluster acoustically and semantically similar environments into the same embedding, which is crucial for generalizing the models for unseen environments. Our result also shows that the CLAP-based text encoder can be highly improved by updating the text encoder. This result demonstrates the importance of minimizing the embedding space gap between the text encoder and the acoustic embedding. 

We plan to expand this approach to cases where we cannot obtain information on the testing environment. While AST embeddings demonstrate competitive performance for arousal and dominance, they do not show improvements for valence compared to models that explicitly use text embeddings. The CL-based representation can address scenarios where noise information is not provided by introducing its audio encoder. CLAP trains the audio encoder to have a similar representation to the ones from the text encoder, which could be useful for extracting environmental information from the audio. For this reason, we plan to investigate how we can improve the noise-robustness of the SER model with a CLAP encoder.  We also plan to investigate the alternative approach of leveraging the inferred caption from an \emph{automated audio captioning} (AAC) model \cite{Mei_2022}, using it as an environmental descriptor

\bibliographystyle{IEEEtran}
\bibliography{refs}
\end{document}